\documentclass[final,5p,times,twocolumn]{elsarticle}

\usepackage{lineno,hyperref}
\journal{Journal of \LaTeX\ Templates}

\usepackage{amsmath,amssymb,hyperref,array,graphicx,wrapfig,caption}
\usepackage{siunitx}
\usepackage{caption}
\usepackage{subcaption}

\newcolumntype{C}[1]{>{\let\newline\\\arraybackslash\hspace{0pt}}m{#1}}
\newcolumntype{L}[1]{>{\raggedright\let\newline\\\arraybackslash\hspace{0pt}}m{#1}}
\newcommand{\mSR}{\textmu SR}

\begin{document}

\begin{frontmatter}

\title{New Frontiers in Muon-Spin Spectroscopy Using Si-Pixel Detectors}

% \address{Radarweg 29, Amsterdam}

% %% or include affiliations in footnotes:
% \author[mymainaddress,mysecondaryaddress]{Elsevier Inc}

% \author[mysecondaryaddress]{Global Customer Service\corref{mycorrespondingauthor}}
% \cortext[mycorrespondingauthor]{Corresponding author}

\author[PI]{Heiko Augustin}
\author[MZ]{Niklaus Berger}
\author[PSI]{Andrin Doll}
\author[PSI,UZ]{Pascal Isenring}
\author[ETH]{Marius Köppel \fnref{note}}
\author[PSI]{Jonas A. Krieger}
\author[PSI]{Hubertus Luetkens}
\author[PI]{Lukas Mandok}
\author[PSI]{Thomas Prokscha}
\author[PI]{Thomas Rudzki}
\author[PI]{André Schöning}
\author[PSI]{Zaher Salman}

\address[PI]{Physics Institute, Heidelberg University, Im Neuenheimer Feld 226, 69120 Heidelberg, Germany}
\address[MZ]{Institute for Nuclear Physics, Johannes Gutenberg-Universität Mainz, Johann-Joachim-Becher-Weg 45, 55099 Mainz, Germany}
\address[PSI]{PSI Center for Neutron and Muon Sciences, 5232 Villigen PSI, Switzerland}
\address[UZ]{Physik Institut, University of Zürich, Winterthurerstrasse 190, CH-8057 Zürich}
\address[ETH]{Institute for Particle Physics and Astrophysics, ETH Zürich, CH-8093 Zürich, Switzerland}
\fntext[note]{Correspondnig author: mkoepp[at]phys.ethz.ch}

\date{\today}

\begin{abstract}
The study of novel quantum materials relies on muon-spin rotation, relaxation, or resonance (\mSR) measurements.
Yet, a fundamental limitation persists: many of these materials can only be synthesized in extremely small quantities, often at sub-millimeter scales.
While \mSR ~offers unique insights into electronic and magnetic properties, existing spectrometers lack a sub-millimeter spatial resolution and the possibility of triggerless pump-probe data acquisition, which would enable more advanced measurements.
The General Purpose Surface-muon instrument (GPS) at the Paul Scherrer Institute (PSI) is currently limited to a muon stopping rate of \SI{40}{\kilo\hertz} to \SI{120}{\kilo\hertz}, a constraint that will become more pressing with the upcoming High-Intensity Muon Beam (HIMB) project.
To overcome these challenges, we demonstrate the feasibility of employing ultra-thin monolithic Si-pixel detectors to reconstruct the stopping position of muons within the sample, thereby significantly enhancing the capability of measuring at higher muon rate.
Additionally, we explore the first steps toward a triggerless pump-probe \mSR ~measurement scheme.
Unlike conventional pump-probe techniques that require external triggers, a triggerless readout system can continuously integrate stimuli pulses into the data stream, allowing real-time tracking of ultra-fast dynamics in quantum materials.
This approach will enable the study of transient states, spin dynamics, and quantum coherence under external stimuli.
\end{abstract}


\end{frontmatter}

\section{Introduction}
The ability to perform muon-spin rotation, relaxation, or resonance (\mSR) measurements on sub-millimeter samples has long been an aspiration in material science, particularly for quantum materials that are challenging to synthesize in large quantities.
Despite current \mSR\ spectrometers, such as the General Purpose Surface-muon instrument (GPS) at the Paul Scherrer Institute (PSI) \cite{gps}, at continuous muon sources providing high-resolution time-domain measurements, their data collection rates (for a typical \SI{10}{\micro\second} data window) have remained limited to \SI{40}{\kilo\hertz} since the invention of the technique.
These spectrometers typically use a limited number of plastic scintillating detectors.
A thin detector is positioned upstream to register muons entering the sample region, while an array of detectors surrounding the sample detects the emitted decay positrons.
A standard spectrometer setup consists of detector pairs (e.g., forward–backward and left–right) along with veto detectors to filter out events from muons that miss the sample.
Although the detector array provides a large solid-angle coverage, it lacks spatial resolution beyond basic directional information.

In general, a \mSR ~measurement starts by detecting an individual incoming spin-polarized muon, triggering a timing clock upon entry.
The clock stops when a positron is detected.
By accumulating a large number of muon and positron events, the temporal evolution of the asymmetry in detector counts on either side of the sample constitutes the \mSR ~signal.
This signal is directly proportional to the time-dependent muon spin polarization along the corresponding axis.
However, this design has two key limitations in existing spectrometers:
\begin{enumerate}[a)]
    \item The origin of the emitted positron is not measured directly and is therefore only assumed to originate from the sample.
    \item To avoid ambiguities from overlapping events, only one muon should be present in the sample at any given time and a single positron should be detected within the measurement window.
\end{enumerate}

Given the \SI{\sim 2.2}{\micro s} lifetime of the muon, a typical \SI{10}{\micro s} time window is required before allowing the next muon to enter the setup.
At PSI, quasi-continuous muon beams are used, meaning that even when the beam rate is restricted, occasional instances of multiple muons entering the sample simultaneously (so-called pile-up) occur.
To mitigate this, a pile-up correction is applied during data analysis: if a new incoming muon is detected before the decay positron from the previous muon is observed, the event is discarded.
Furthermore, all events in which a second positron is detected during the same data gate of \SI{10}{\micro s} are also removed from the analysis.
As a result, data acquisition times often extend to several hours per measurement.

To improve on this approach, several attempts have been made.
Chow and colleagues~\cite{chow2003multi} developed a scheme to enable time-differential \mSR\ measurements at higher rates at TRIUMF.
The scheme involved sandwiching a thin sample between two segmented forward and backward scintillating counters.
Each counter, segmented into \qtyproduct{4 x 4}{mm}, was used to tag incoming muons and emitted positrons, effectively generating separate data sets from opposite forward–backward segments, each theoretically capable of running at rates up to \SI{40}{\kilo\hertz}.
However, this approach had several technical challenges.
The scintillators and light guides used to transmit the signals were placed inside the cryostat alongside the sample, making the setup highly complex and fragile.
This design also led to the degradation of the scintillators and light guides, with cracks developing due to thermal cycling.
Additionally, the available electronics at the time could not handle the high data rates generated by the system.
The setup also suffered from cross-talk issues between different segments.

Another approach to overcome current limitations involves using scintillating fibres to achieve spatial resolution.
This concept was first proposed in 2006~\cite{shiroka2006position}, and a recent study at TRIUMF successfully demonstrated its feasibility of tracking positrons~\cite{sugisaki2024development}.
Although the fibres in this setup were able to measure a \mSR~signal, the achieved muon rate of only \SI{800}{muons / s} and a spatial resolution worse than one millimeter were far from meeting the sub-millimeter requirement for a potential spectrometer at high-rate muons sources.
Furthermore, this detector was used exclusively for positron tracking, as the material budget of the fibre detector was too high to allow for precise muon tracking.

Recent advancements in silicon pixel detector technology, particularly with the MuPix11 chip~\cite{mupix10,heikoPhd}, have enabled high-precision particle tracking capabilities at low momenta.
These monolithic silicon pixel detectors achieve a spatial resolution of approximately \SI{23}{\micro m} per hit and a time resolution below \SI{15}{\nano\second} within a \SI{50}{\micro m} silicon layer.
Integrating these detectors into \mSR ~spectrometers presents a promising path to overcome the existing lateral resolution and rate constraints.

Furthermore, triggerless readout systems are becoming increasingly relevant in particle physics experiments due to their ability to handle high data rates and to enable online event reconstruction.
However, to date, \mSR\ measurement at pulsed muon facilities such as ISIS and J-PARC has been conducted using a triggered readout system. At continuous muon sources, a limited triggerless data acquisition (DAQ) system has been implemented, where a good event is triggered by an incoming muon.
One of the most transformative applications of an advanced triggerless approach is its potential to revolutionize pump-probe \mSR\ measurements.
Traditional pump-probe experiments rely on an external trigger, such as a laser pulse, to initiate data collection.
This method imposes limitations on temporal resolution and event selection in currently used \mSR\ DAQ.
A pump-probe triggerless readout system, on the other hand, can integrate the timing of laser pulses directly into the data stream, allowing for continuous tracking of system dynamics without predefined timing windows.
This advancement would enable the direct observation of ultra-fast processes and transient states, making it a novel tool for studying quantum coherence, spin dynamics, and nonequilibrium phenomena in materials subjected to optical, microwave, or pressure-driven excitation.
In this work, we present the first steps towards such a system by demonstrating the feasibility of conducting \mSR\ measurements with an advanced triggerless readout.

\section{Spectrometer Design \& Simulation Studies}

A potential spectrometer could consist of two ultra-thin silicon pixel detector layers arranged in a box-like configuration around the sample.
This design allows for large spatial coverage, and one could perform vertex reconstruction to significantly improve the spatial resolution.
Furthermore, the ability to correlate incoming muons with outgoing positrons enables rejection of background events and an increase in usable muon rate.
For an initial demonstrator, we developed a quad module~\cite{masterLukas} (shown in Figure~\ref{quad-pic} (a) and (b)) that incorporates four MuPix11 chips~\cite{masterLukas}.
Four such modules were used in all presented test configurations, with two upstream and two downstream layers relative to the sample under investigation.

\begin{figure*}[h]
    \centering
    \includegraphics[width=0.8\textwidth]{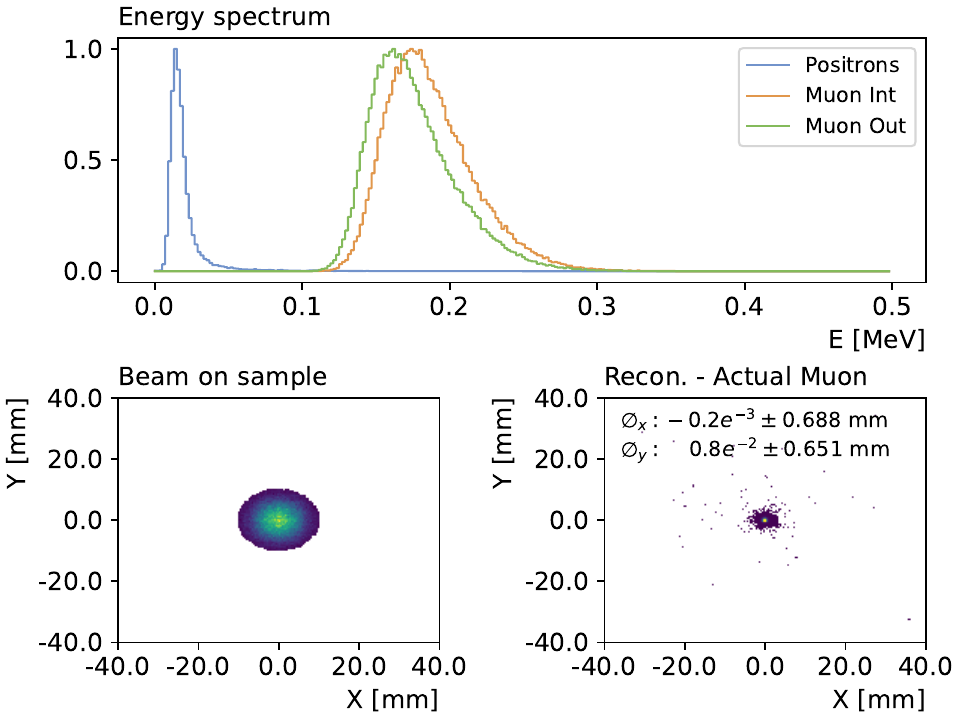}
    \caption{Simulation Results: Top: Deposited energy in the Si chip for positrons (blue), muons in the inner layers (orange) and muons in the outer layers (green). Bottom left: Muons stopped in the sample. Bottom right: Difference between the reconstructed and actual Muon position from the Monte-Carlo simulation.}
    \label{fig:musr-sim}
\end{figure*}
To optimize the detector design and evaluate potential uncertainties of our prototype detector we conducted initial simulations based on the \SI{28}{MeV/c} muon beam used at the GPS instrument.
We positioned two layers of Si-chips upstream and downstream.
To estimate the uncertainty in vertex reconstruction, we used MusrSim simulations~\cite{musr-sim,musr-sim2}, which are based on GEANT4 \cite{geant4}, to model the scattering behaviour of muons passing through \SI{50}{\micro m} of silicon.
Figure~\ref{fig:musr-sim} presents the simulation results for a setup where the inner Si-chips are positioned at a distance of $r=$\SI{10}{mm} from the sample.
Our findings indicate that the primary source of uncertainty in vertex reconstruction for an incoming muon trajectory -- determined from hits on the upstream layers -- is primarily due to multiple-Coulomb scattering occurring within the inner Si-chip.

The cross-section of the muon beam on the sample is shown in the lower-left part of Figure~\ref{fig:musr-sim}.
In the lower-right part of the same figure, the difference between the reconstructed and actual muon positions at the target is presented.
The standard deviation of this difference serves as a measure of the uncertainty in determining the muon's lateral position on the sample via vertex reconstruction.
In the proposed configuration, the estimated uncertainty is less than~\SI{0.7}{mm}.

One way to further improve the resolution is by combining the extrapolated tracks of the muon and the coincident positron to establish a match.
The key to efficient track extrapolation for both muons and positrons lies in the ability to accurately distinguish between these two particle species.
An efficient approach to achieve this differentiation is to optimize the Time-over-Threshold (ToT) stopping power measurement implemented in most novel Si-Pixel chips, based on the momentum distribution of the incoming \SI{28}{MeV/c} muons and the decay positrons.
The top of Figure \ref{fig:musr-sim} presents the simulated energy deposition spectra for positrons (blue), muons in the inner layer (orange), and muons in the outer layer (green).
One can clearly see that the muons deposit much higher energy than the positrons, providing a basis for this differentiation.

\section{Triggerless Readout System}

The MuPix11 chip is divided into three sub-matrices, each transmitting zero-suppressed, time-unsorted hit data at a rate of \SI{1.25}{Gbit/s}.
The data is streamed continuously from the detector without an external trigger, ensuring a high-rate, real-time data flow.
Synchronization across all chips is managed by a \SI{125}{MHz} clock, provided by a custom FPGA-based front-end board (FEB).
This system is derived from the Mu3e experiment~\cite{mu3edaq,ieeeDataFlow} and adapted to the requirements of \mSR.
Each FEB supports 45 high-speed readout channels operating at \SI{1.25}{Gbit/s}.
Two of the prototype quad modules are connected to a single FEB via specialized detector adapter boards (DABs) that incorporate onboard equalizers for signal integrity.
One FEB reads the first two layers upstream of the sample, while the second reads the two layers downstream.
The FEBs sort the data in time using onboard RAM, where timestamps act as addresses, ensuring sequential readout.

The sorted data is then transmitted optically to a Terasic DE5a-Net-DDR4 board~\cite{de5net} equipped with an Arria10 FPGA.
Using a custom Direct Memory Access (DMA) engine, hit data from all four layers is transferred to the main RAM of a PC at a peak rate of \SI{38}{Gbit/s}~\cite{mariusMaster,gpu_bruch}.
Data acquisition and chip configuration are managed via the Maximum Integration Data Acquisition System (MIDAS)~\cite{midas1,midas2}.
Since no external trigger is employed, data is read in continuous chunks, dynamically adjusting to storage availability to prevent bottlenecks.

\section{First Results}

\begin{figure*}[h]
    \includegraphics[width=\textwidth]{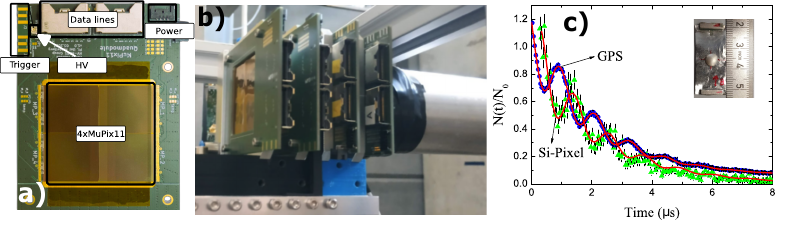}
    \caption{(a) A photo of a quad hosting four Si-pixel chips. (b)~The arrangement of four tracking layers using quads modules in the prototype spectrometer. (c)~Comparison of the $\mu$SR signal measured by the Si-Pixel spectrometer vs. the GPS reference of a 6~mm diameter Al disc placed between two permanent magnets producing a transverse field of \SI{6.3}{mT} at the centre, taken from \cite{mandok2025advancedmuonspinspectroscopyhigh}.}
    \label{quad-pic}
\end{figure*}

Here, the results published in another work \cite{mandok2025advancedmuonspinspectroscopyhigh} are summarized.
In contrast to the simulation, where we only used \SI{50}{\micro m} thick MuPix11 chips, we used one chip with \SI{50}{\micro m} in front of the sample, while the remaining chips had thicknesses between \SI{70}{\micro m} and \SI{100}{\micro m}.
To demonstrate the performance of the prototype, a \SI{6}{mm} diameter aluminium disc placed between two permanent magnets, generating a transverse field of \SI{6.3}{mT} was used as a sample.
Figure \ref{quad-pic} (c) illustrates the time difference between the incoming muon and the outgoing positron, utilising only the upstream layers.
For comparison, Figure \ref{quad-pic} (c) also includes reference data from the General Purpose Surface-muon instrument (GPS) \cite{gps} measuring the very same sample.
In both measurements, the Larmor frequency (Si-Pixel:~\SI{0.868\pm0.057}{MHz}, GPS:~\SI{0.857\pm0.002}{MHz}) and the Lorentzian width (Si-Pixel:~\SI{0.18\pm0.055}{1/\micro s}, GPS:~\SI{0.176\pm0.01}{1/\micro s}) were found to be nearly identical within the error margin.
Notably, the uncorrelated background in the Si-Pixel spectrum was nearly zero, an improvement over GPS due to the ability to reconstruct positrons and muons sharing the same vertex and therefore eliminating any uncorrelated background.
The phase difference between the measurements is due to the different orientations of the detectors relative to the initial polarisation of the muons.
The GPS apparatus was placed -60 degrees, while the Si-Pixel detector was oriented 90 degrees relative to the spin orientation of the muons.

\section{Conclusion and Future Outlook}
In this work, we have demonstrated the feasibility of integrating Si-pixel detectors into a \mSR\ spectrometer to achieve precise vertex reconstruction, enabling high-rate measurements on sub-millimeter samples.
The implementation of a triggerless readout system allows for efficient data acquisition, overcoming the limitations of traditional spectrometers and significantly enhancing the measurable muon rate.

One of the most transformative applications of such a spectrometer is the ability to perform pump-probe \mSR\ measurements in a triggerless mode and with a high time resolution.
Conventional pump-probe spectrometers rely on an external trigger, such as a laser pulse, to initiate data collection.
However, with the proposed triggerless readout system that can track laser pulses in real time, it becomes possible to capture the full dynamics of a system before, during and after the external stimulus.

This capability is critical for studying quantum coherence, spin dynamics, and ultrafast phenomena in materials where external stimuli -- such as optical, microwave, or pressure pulses -- induce transient states.
By synchronizing muon detection with the external excitations, the proposed scheme paves the way for enhancing the power \mSR\ its applications for quantum material research, pushing the technique into a new era of time-resolved studies.
Future work will focus on optimizing the synchronization of external stimuli with muon detection and further enhancing the spatial and temporal resolution to meet the demanding requirements of next-generation quantum material investigations.

\section{Acknowledgments}
This research is funded by the Swiss National Science Foundation (SNF-Grant No. 200021\_215167).
All experiments were performed at the Swiss Muon Source SµS, Paul Scherrer Institute, Villigen, Switzerland. The authors are grateful to Frank Meier, Andreas Knecht and Alex Amato for fruitful discussions. We also thank Martin Müller for his initial help porting the Mu3e DAQ system.

\end{document}